\documentclass[%
 reprint,
 amsmath,amssymb,
 aps,
prb,
floatfix,
]{revtex4-2}

\usepackage{graphicx}
\usepackage{dcolumn}
\usepackage{xcolor}
\usepackage{hyperref}
\usepackage{bm}

\hypersetup{
  colorlinks=true,
  linkcolor=blue,  
  citecolor=blue,  
  urlcolor=blue    
}

\newcommand{\dyadA}{\widehat{ \alpha }}

\begin{document}

\preprint{APS/123-QED}

\title{Rayleigh Bound States in the Continuum}

\author{Ilya Karavaev$^{1,2}$}

\author{Mingzhao Song$^{1}$}

\author{Andrey Bogdanov$^{1,2}$}
\email{Corresponding author: a.bogdanov@hrbeu.edu.cn}

\affiliation{$^{1}$ Qingdao Innovation and Development Center, Harbin Engineering University, Qingdao 266404, China}

\affiliation{$^{2}$ Faculty of Physics and Engineering, ITMO University, Saint Petersburg 197101, Russia}


\date{\today}

\begin{abstract}
We predict a class of bound states in the continuum (BICs) in non-subwavelength periodic metasurfaces---Rayleigh BICs---that emerge precisely at Rayleigh anomalies, where diffraction channels open. In contrast to conventional symmetry-protected and accidental BICs, which are typically engineered in the subwavelength regime with only a few radiation channels, Rayleigh BICs remain nonradiating far beyond this limit, even when multiple diffraction channels are open. Their formation originates from the interplay between collective lattice resonances, Rayleigh anomalies, and anapole states of the constituent metaatoms. We show that Rayleigh BICs are characterized by an unusual cubic scaling of the quality factor in reciprocal space, ($Q \propto 1/k^3$). Our results establish a constructive route for engineering nonradiating states in metasurfaces beyond the subwavelength regime.
\end{abstract}

\maketitle



Bound states in the continuum (BICs) are nonradiating eigenstates embedded in the spectrum of propagating waves in the surrounding space~\cite{Hsu16,koshelev2023bound,azzamPhotonicBoundStates2021}. In photonics, they provide a mechanism for suppressing radiative losses and realizing resonances with, in principle, unbounded quality factors~\cite{chenObservationMiniaturizedBound2022b}. This has established BICs as a key concept in metasurfaces~\cite{hsuObservationTrappedLight2013}, diffraction gratings~\cite{chang-hasnainHighcontrastGratingsIntegrated2012}, and photonic crystals~\cite{koshelevMetaopticsBoundStates2019}, with applications in enhanced light-matter interaction, nonlinear optics, sensing, and wavefront control~\cite{kangApplicationsBoundStates2023}.

In a {\it subwavelength} metasurface, the radiative continuum comprises only four outgoing channels: upward and downward radiation, each with two orthogonal polarizations. Coupling to these channels can vanish either due to the geometrical symmetry of the unit cell or through fine tuning of structural or material parameters~\cite{zhenTopologicalNatureOptical2014}. This naturally leads to the conventional distinction between symmetry-protected and accidental BICs~\cite{koshelev2023bound}. The small number of available radiation channels makes the subwavelength regime the natural setting for most established BIC design strategies in periodic photonic structures~\cite{wangOpticalBoundStates2024}.

Beyond the subwavelength regime, the opening of additional diffraction channels generally destroys conventional symmetry-protected and accidental BICs~\cite{sadrievaTransitionOpticalBound2017,yangAnalyticalPerspectiveBound2014}. Removing this restriction is important both fundamentally and practically, since it relaxes the stringent design constraints imposed by deeply subwavelength periodicity.  Existing routes rely either on engineering the radiative environment~\cite{cerjanObservationBoundStates2021} or on varying geometrical parameters to cancel leakage into higher-order diffraction channels~\cite{wangHybridBoundStates2026,yeSingularPointsPolarizations2020,bulgakovLightTrappingAbove2015}. However, the number of independent tuning parameters required to suppress radiation grows with the number of open channels, making this approach increasingly impractical~\cite{bykovAlgebraicApproach2024}. Therefore, a general practical recipe for overcoming this challenging problem and engineering BICs in non-subwavelength metasurfaces is still lacking.

In this work, we demonstrate that robust BICs can exist in non-subwavelength metasurfaces, i.e., beyond the diffraction limit. Using multiple-scattering theory, we show that a distinct class of BICs emerges when a {\it lattice-resonance} (LR) crosses the Rayleigh anomalies (RAs) corresponding to opening diffraction channels. We refer to these states as {\it Rayleigh BICs} as their formation originates from the combined effect of collective LR and {\it anapole} response of an individual scatterer. Although the anapole of an isolated metaatom is not an eigenmode, the revealed case is a beautiful exception. We show that the anapole state in a periodic array turns into a true nonradiating eigenstate at the RA. At the anapole frequency, the metaatom does not radiate in any direction, so coupling to all open diffraction channels vanishes simultaneously even if multiple diffraction channels are open, allowing the BIC to persist beyond the subwavelength regime. We first consider a dipole lattice and then extend the model to arbitrary high multipoles, showing that Rayleigh BICs can be realized in arrays of finite-sized metaatoms by spectrally aligning the anapoles corresponding to different multipolar scattering channels.

\begin{figure}[t!]
\includegraphics[width=1\linewidth]{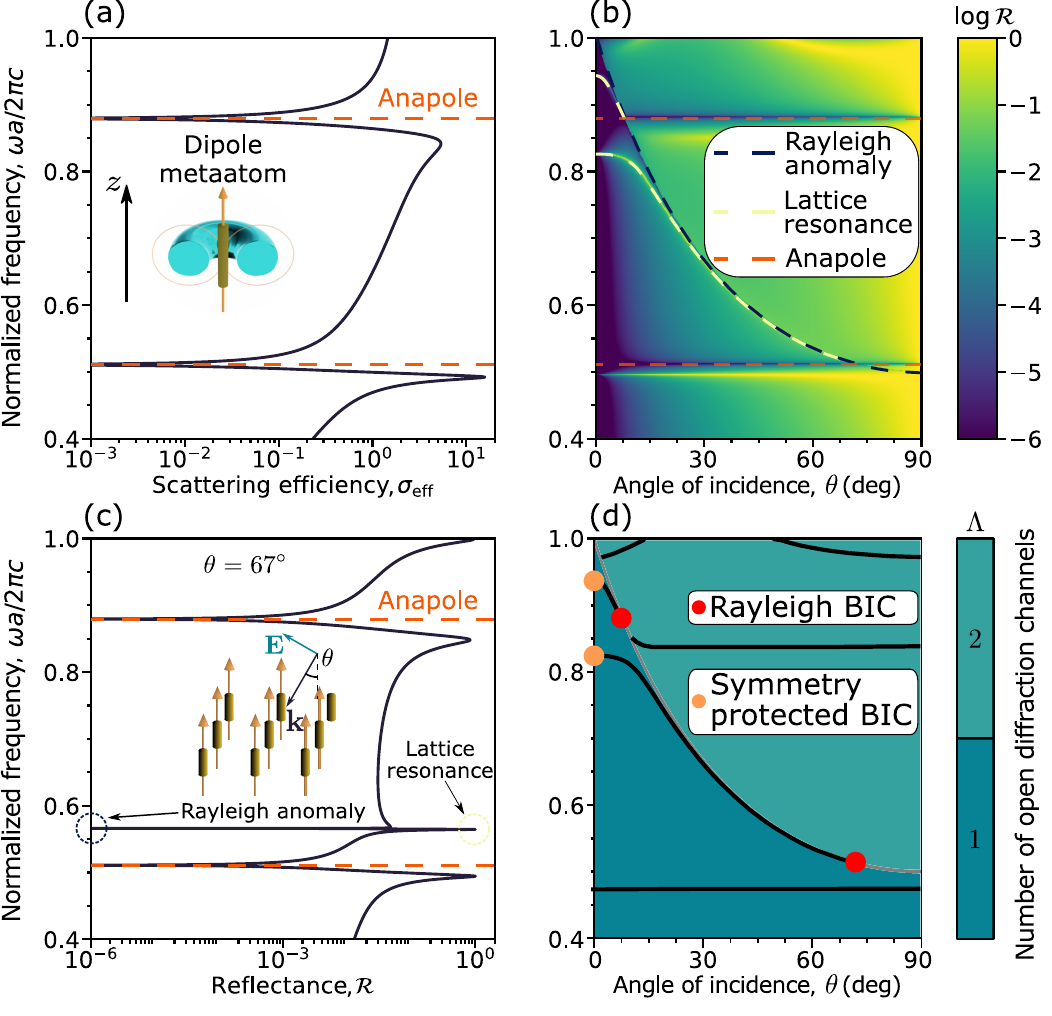}
\caption{\label{fig:2} Optical response in the framework of the dipole approximation. Panel (a) shows the spectrum of scattering efficiency for an individual particle, panel (b) demonstrates the reflection coefficient spectrum at all angles of incidence for a square lattice built with the aforementioned scatterers, and panel (c) displays the cross-section of this map at $\theta = 67^\circ$. Panel (d) reveals BIC positions and resonance curves with a different number of open diffraction channels.}
\end{figure}

We start from a toy model of a square lattice with period $a$, formed by lossless point electric dipoles placed in air [see the inset in Fig.~\ref{fig:2}(c)]. Without loss of generality, we restrict our analysis to vertical $z$-oriented dipoles, since along the $\Gamma \mathrm{X}$ direction the vertical and in-plane dipole components transform according to different irreducible representations of the $C_{2v}$ point group and, therefore, do not hybridize (see the Supplemental Material)~\cite{sadrievaMultipolarOriginBound2019}. We thus describe the particles by the polarizability tensor $\dyadA=\mathrm{diag}(0,0,\alpha_e)$, where $\alpha_e=\mathrm{i}3a_1/(2k_0^3)$ and $a_1$ is the electric-dipole Mie scattering coefficient~\cite{bohren2008absorption}. For concreteness, the Mie polarizability is taken to correspond to a high-index dielectric sphere with permittivity $\varepsilon=50$ and radius $R=0.2a$. Figure~\ref{fig:2}(a) shows the scattering efficiency of a single dipole particle on a logarithmic scale. The spectrum exhibits two Mie resonances at $\omega a/2\pi c \approx 0.49$ and $\omega a/2\pi c \approx 0.84$, together with two pronounced scattering dips at $\omega a/2\pi c \approx 0.51$ and $\omega a/2\pi c \approx 0.88$. These dips correspond to anapole states, when the corresponding Mie coefficient vanishes, i.e., $a_1=0$~\cite{miroshnichenkoNonradiatingAnapoleModes2015,koshelevNonradiatingPhotonicsResonant2019}.

Within multiple-scattering theory, the collective response of the coupled-dipole array is fully described by the effective, or dressed, polarizability~\cite{babichevaMultipoleLatticeEffects2021, utyushevCollectiveLatticeResonances2021}
\begin{equation}
\alpha_\text{eff}(\omega,k_\|)=\frac{\alpha_e}{1-\alpha_e S(\omega,k_\|)}.
\label{eq:alpha_eff}
\end{equation}
Here $S(\omega,k_\|)$ is the lattice sum~\cite{celepcikay2008choosing, Supplement}, $k_\|=k_0\sin\theta$ is the in-plane Bloch wavenumber, $k_0=\omega/c$, and $\theta$ is the angle of incidence. The amplitude reflection coefficient for a TM-polarized wave incident at angle $\theta$ corresponding to the specular reflection  is given by~\cite{Supplement, allayarovAnalyticalModelMetasurfaces2025}:
\begin{equation}
r=-\frac{2\pi i k_0}{a^2}\frac{\sin^2\theta}{\cos\theta}\,\alpha_\text{eff}.
\label{eq:refl}
\end{equation}
The energy reflection coefficient is $\mathcal{R}(\omega,\theta)=|r|^2$. 

Figure~\ref{fig:2}(b) shows reflectance $\mathcal{R}$ as a function of the normalized frequency $\omega a /(2\pi c)$ and angle of incidence $\theta$. Several characteristic features can be identified in this map. (i) The metasurface is fully transparent ($\mathcal{R}=0$) at the anapole frequencies (orange dashed lines) independently of the angle of incidence. The anapole frequency is angle-independent because metaatoms do not interact via scattering at this frequency. (ii) At normal incidence, the structure is transparent because $z$-aligned dipoles cannot be excited by the incident wave. Therefore, the array supports symmetry-protected BICs at the $\Gamma$ point~\cite{gladyshevBoundStatesContinuum2022}. (iii) The dashed black curve indicates the RA~\cite{rayleigh1907dynamical}, while the dashed yellow curves mark the LRs~\cite{zouSilverNanoparticleArray2004, kravets2018plasmonic,utyushevCollectiveLatticeResonances2021}. The reflection spectrum at $\theta=67^\circ$ shows how close the LR is to the RA [Fig.~\ref{fig:2}(c)]. (iv) The LR appears as a pronounced reflection maximum with $\mathcal{R}=1$, whereas $\mathcal{R}=0$ at the nearby RA~\cite{swiecickiSurfacelatticeResonancesTwodimensional2017}.

The eigenmodes of the metasurface are defined by the complex poles of the effective polarizability $\alpha_\text{eff}$ [see Eq.~\eqref{eq:alpha_eff}].  Figure~\ref{fig:2}(d) shows the real parts of the corresponding eigenfrequencies as functions of angle $\theta$. The states at the $\Gamma$-point marked by orange markers are the symmetry-protected BICs. According to previous studies, there are no off-$\Gamma$ BICs in a lattice of vertical dipoles because their radiation pattern vanishes only along the vertical axis~\cite{allayarovAnalyticalModelMetasurfaces2025,gladyshevBoundStatesContinuum2022,abujetasCoupledElectricMagnetic2020,sadrievaMultipolarOriginBound2019,WeiLiu-PRL-2019}. However, our analysis reveals an {\it exception} that was not identified in previous studies. The dispersion of the LR tangentially approaches the RA and touches it exactly at the anapole frequency, where a Rayleigh BIC emerges. The same conclusion follows from Eq.~\eqref{eq:alpha_eff}. The dispersion of the lattice-resonance mode is determined by the pole of the effective polarizability~\cite{Supplement}
\begin{equation}
    S(\omega,k_\|)=\alpha_e^{-1}.
    \label{eq:lattice_res_condition}
\end{equation}
At the RA, the normal wavenumber of the opening diffraction order vanishes, giving rise to a square-root branch point in the scattering matrix~\cite{PhysRevLett.127.277401}. As this threshold is approached, the lattice sum $S(\omega,k_{\parallel})$ diverges~\cite{markelDivergenceDipoleSums2005}. At the same time, $\alpha_e^{-1}$ diverges as the frequency approaches the anapole. The Rayleigh BIC condition agrees with the requirement that the numerator and denominator of the complex reflection coefficient $r$ vanish at the same real frequency~\cite{bezusBoundStatesContinuum2018}. Since $r\propto\alpha_{\mathrm{eff}}$ in Eq.~\eqref{eq:refl}, the single-particle anapole supplies the numerator zero, while the lattice resonance supplies the denominator zero. These zeros coincide at the Rayleigh anomaly.

\begin{figure}[t!]
\includegraphics[width=1\linewidth]{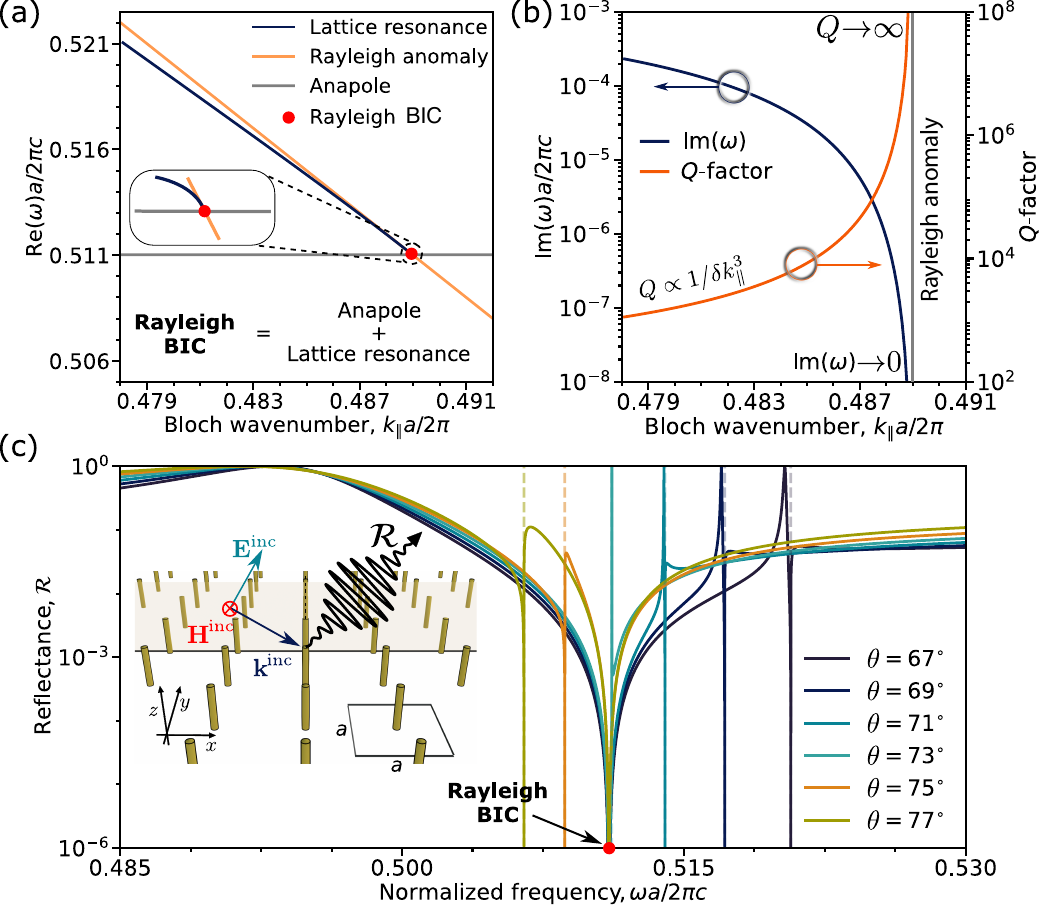}
\caption{Panels (a) and (b) show the dimensionless real and imaginary parts of the LR frequency, respectively, as a function of the in-plane Bloch wavenumber $k_\|$. The red dot indicates the frequency and coordinates in reciprocal space where the Rayleigh BIC is formed. Panel (c) shows the reflection coefficient as a function of the incident wave frequency at fixed angles $\theta$. Dashed lines with the corresponding color indicate the opening of the first diffraction channel.\label{fig:3}}
\end{figure}

Figure~\ref{fig:3}(a) shows in detail how the Rayleigh BIC emerges at the intersection of three spectral features in one point: (i) the RA, (ii) the LR, and (iii) the anapole. The imaginary part of the complex eigenfrequency of the LR, shown in Fig.~\ref{fig:3}(b), tends to zero at this point, while the corresponding $Q$ factor diverges, indicating the formation of a genuine non-radiating state. The Rayleigh BIC is characterized by an unusual cubic scaling of the quality factor $Q\propto |\delta k_\|^{-3}|$, which distinguishes it from both symmetry-protected and accidental BICs~\cite{Supplement}. This scaling is governed by the nonanalytic asymptotics of the lattice Green function near an RA, where the lattice Green function has a branch cut~\cite{PhysRevLett.127.277401,BetzLPR2025,swiecickiSurfacelatticeResonancesTwodimensional2017}. At the RA, the physical dispersion branch ends. Below the anapole frequency, the mode continues onto the Riemann sheet corresponding to the incoming-wave condition~\cite{Jackson-ch7,allardQuantumTheoryPlasmon2021,koenderinkComplexResponsePolariton2006,abdrabou2022frequency}. This mode contributes to the completeness of the modal expansion for the scattering problem without producing a distinct spectral resonance. The full dispersion therefore requires a multisheet Riemann surface~\cite{PhysRevA.99.063818,tamirGuidedComplexWaves1963,koenderinkComplexResponsePolariton2006,bellamine1994guided,Jackson-ch7}.

Figure~\ref{fig:3}(c) shows the reflectance $\mathcal{R}$ as a function of frequency for several incidence angles $\theta$. The colored dashed vertical lines mark the RA for each $\theta$, where the reflectance vanishes, $\mathcal{R}=0$. The reflectance also vanishes at the anapole frequency, which is independent of the angle of incidence. Between these two minima, the reflectance reaches a maximum, $\mathcal{R}=1$, corresponding to the LR. As the angle varies, the Rayleigh-induced reflection minimum shifts across the spectrum and approaches the anapole frequency, while the LR remains sandwiched between the anapole and the RA. As the LR approaches the anapole, its linewidth narrows, indicating an increase in the $Q$ factor. At the critical angle, $\theta\approx73^\circ$, the anapole, the RA, and the LR coincide. At this point, the two reflection minima merge with the reflection maximum. This behavior in reflection resembles, in an inverted form, electromagnetically induced transparency~\cite{lukin2001controlling}. For $\theta>73^\circ$, the lattice-resonance branch continues onto a different Riemann sheet, where it no longer produces a distinct reflectance peak.

Symmetry-protected and accidental BICs can be understood within the multipolar picture of radiation from the unit cell~\cite{sadrievaMultipolarOriginBound2019,WeiLiu-PRL-2019,allayarov2024}. In this framework, a BIC appears when the directions of all open diffraction channels coincide with zeros of the directivity diagram of the polarization currents in the unit cell. In the subwavelength regime, this condition can be achieved either by symmetry or through interference of several multipoles. Beyond the subwavelength regime, however, additional diffraction orders become open, and each of them introduces a new radiation channel that must be suppressed. This requires increasingly fine control of the directivity diagram of the unit cell and, correspondingly, a growing number of tuning parameters~\cite{bykovAlgebraicApproach2024}. As a result, BICs in the presence of multiple open diffraction channels become unstable and substantially more difficult to engineer. In contrast, the Rayleigh BIC can persist beyond the subwavelength regime, even when multiple diffraction channels are open, because the anapole condition suppresses radiation from each metaatom into all radiative channels simultaneously. In this sense, the Rayleigh BIC provides a robust mechanism for BIC formation in non-subwavelength metasurfaces.

\begin{figure}[t]
\includegraphics[width=1\linewidth]{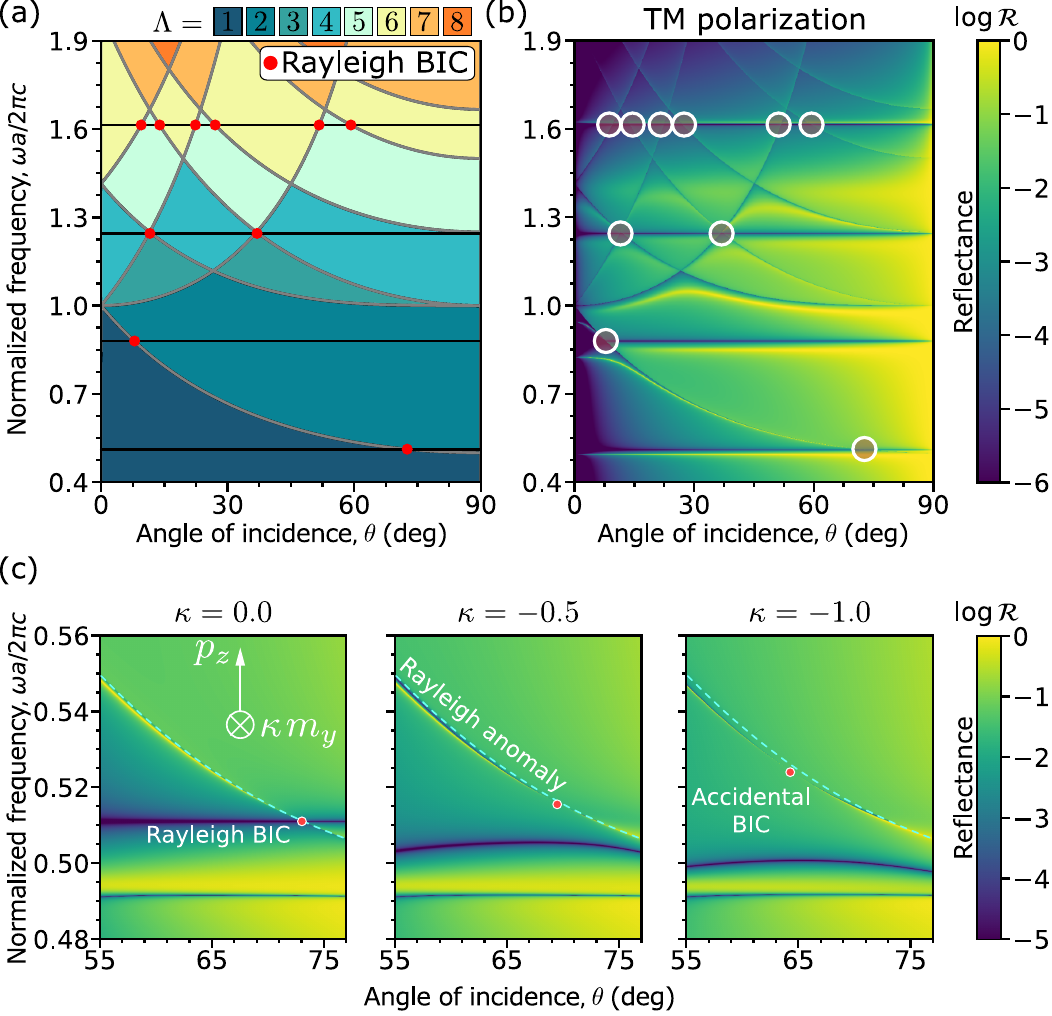}
\caption{\label{fig:4}  (a) Dispersion of the RAs (gray solid lines). $\Lambda$ encodes the number of open diffraction channels. Black horizontal lines denote anapole frequencies. Red dots at the intersections of anapole lines and RAs mark the Rayleigh BICs. (b) Reflectance map for TM polarization, with white circles marking the Rayleigh-BIC points. Panel~(c) shows the transition from a Rayleigh BIC to an accidental BIC upon inclusion of the magnetic-dipole contribution with relative weight $\kappa=0, 0.2, 1$, as illustrated in the inset.}
\end{figure}

Figures~\ref{fig:4}(a) and \ref{fig:4}(b) illustrate the robustness of Rayleigh BICs when multiple diffraction channels are open. Panel~(a) shows the RAs (gray solid lines) together with the number of open diffraction channels $\Lambda$. The black horizontal lines mark the anapole frequencies. The intersection points of RAs with anapoles marked by red dots correspond to the Rayleigh BICs. Importantly, these points occur not only in the subwavelength regime but also in regions where multiple diffraction channels are open. Panel~(b) shows the corresponding reflectance map for TM polarization. The white circles mark the Rayleigh BICs, where the lattice mode linewidth shrinks, indicating a diverging Q factor.

We further analyze the role of other multipoles beyond the electric dipole approximation. The vertical electric dipole $p_z$ and horizontal magnetic dipole $m_y$ belong to the same irreducible representation and can therefore contribute to the same mode~\cite{sadrievaMultipolarOriginBound2019,Supplement}. A lattice model with electric and magnetic dipoles captures both accidental and symmetry-protected BICs in all-dielectric metasurfaces~\cite{abujetasCoupledElectricMagnetic2020,abujetasTailoringAccidentalDouble2022,gladyshevInverseDesignAlldielectric2023,allayarov2024}. Figure~\ref{fig:4}(c) shows the reflectance of a lattice of vertical electric dipoles $p_z$ for different relative contributions of the horizontal magnetic dipole $m_y$, set by the weight coefficient $\kappa$~\cite{Supplement}. The magnetic polarizability is $\alpha_m=3\mathrm{i}b_1/(2k_0^3)$, with both Mie coefficients $a_1$ and $b_1$ evaluated for a sphere with $\varepsilon=50$ and $R=0.2a$. As $\kappa$ increases, the Rayleigh BIC evolves continuously into an accidental BIC. This transformation changes the divergence asymptotics of the $Q$ factor from $Q\propto|\delta k_\||^{-3}$ near the Rayleigh BIC to $Q\propto|\delta k_\||^{-2}$ near the accidental BIC~\cite{Supplement}. Reversing the sign of $\kappa$ shifts the Rayleigh BIC  onto the second Riemann sheet, where it no longer produces distinct resonance features~\cite{Supplement}.

Finally, we show that Rayleigh BICs can exist beyond the dipole approximation, even when multiple diffraction channels open. The central idea is to compose a metasurface from metaatoms whose Mie coefficients in different multipolar channels vanish at the same frequency. This effect is known as a \textit{hybrid anapole}~\cite{lukyanchuk2017,valero2021}. It is a counterpart to superscattering, which occurs when Mie resonances of different multipoles are spectrally aligned~\cite{ruan2010}. To illustrate this approach, we design a spherical dielectric metaatom with a core and five concentric shells. Its electric and magnetic Mie coefficients vanish simultaneously up to octupole order, $a_\ell=b_\ell=0$ for $\ell=1,2,3$, at $k_0R=1$, where $R$ is the sphere's outer radius. Figure~\ref{fig:5}(a) shows the total and partial scattering cross sections. We detail the sphere design in~\cite{Supplement}.

We arrange these spheres in a square lattice with $a/R=7$. Figure~\ref{fig:5}(b) shows the TM reflectance map calculated with MULTEM~3~\cite{multem2024}, retaining electric and magnetic multipoles through $\ell=8$. The map reveals two arc-shaped resonant branches, labeled A and B. Each mode has two open diffraction channels ($\Lambda=2$) into the upper and lower half-spaces. Both resonances narrow as they approach the RAs and become unresolved in the reflectance map near the intersections with the anapole frequency, where they turn into Rayleigh BICs.

\begin{figure}[t]
\includegraphics[width=1\linewidth]{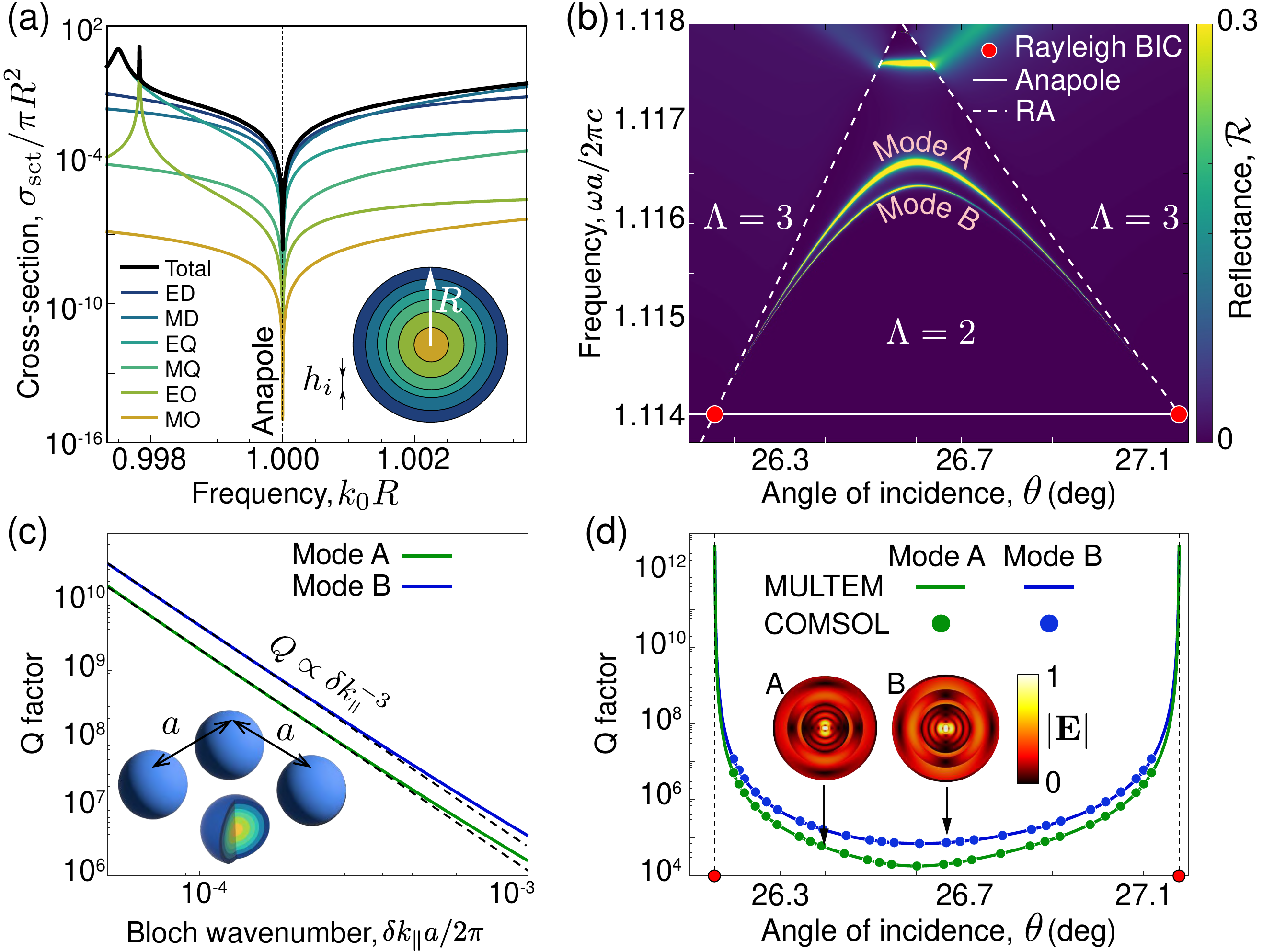}
\caption{\label{fig:5} (a) Total and partial scattering cross-sections of the multilayer spherical metaatom near the anapole. The design of the metaatom shown in the inset is described in the Supplemental Material.  (b) The reflectance map from the metasurface composed of core-shell metaatoms calculated by MULTEM~3 for a TM-polarized incident wave near RAs. (c) Asymptotics of the Q factor near the Rayleigh BIC at $\theta=26.16^\circ$. The dashed lines show the cubic asymptotics, $Q\propto \delta k_\parallel^{-3}$. (d) Radiative Q factor for modes A and B calculated numerically in MULTEM~3 (lines) and COMSOL (dots). Insets show the eigenmode profiles at $\theta=26.6^\circ$.}
\end{figure}

Figure~\ref{fig:5}(d) compares the radiative $Q$ factors obtained from the COMSOL eigenfrequency solver~\cite{comsol} and from the complex pole analysis of the MULTEM~3 multiple-scattering matrix, calculated through $\ell=8$~\cite{multem2024}. The two methods agree perfectly, however, near the Rayleigh anomalies, COMSOL convergence requires placing the perfectly matched layers farther from the array, which increases the computational cost. The MULTEM~3 pole calculation resolves radiative $Q$ factors above $10^{12}$, close to the limit set by the numerical precision of our method. The insets show the electric-field profiles of modes A and B near $\theta=26.6^\circ$. To confirm additionally that the observed states are Rayleigh BICs, we analyze the asymptotic behavior of $Q$ near the anapole frequency. Figure~\ref{fig:5}(c) shows that, near the Rayleigh-anapole crossing at $\theta=26.16^\circ$, both modes inherit the inverse-cubic scaling predicted by the dipole model, i.e., $Q\propto \delta k_\|^{-3}$.


The Rayleigh BIC can be viewed as the ultimate nonlocal limit of a surface LR, in which the array's collective response becomes singular at the opening of a diffraction channel~\cite{PhysRevLett.133.053801,10.1088/1361-6633/ae8311}. Surface LRs, RAs, and Wood anomalies have been extensively studied~\cite{hessel1965new,zouSilverNanoparticleArray2004,auguieDiffractiveArraysGold2010,humphreyPlasmonicSurfaceLattice2014,swiecickiSurfacelatticeResonancesTwodimensional2017,kravets2018plasmonic,utyushevCollectiveLatticeResonances2021,babichevaMultipoleLatticeEffects2021,markelDivergenceDipoleSums2005}. These studies, however, do not provide a unified theory linking these phenomena to BIC formation at a diffraction threshold. The coexistence of BICs with surface LRs has also been reported in plasmonic lattices~\cite{trinhCoexistenceSurfaceLattice2022}. Reference~\cite{monticone2017bound} discussed a related Rayleigh-anomaly mechanism, interpreting trapped states in gratings as the collision of a leaky pole with a Rayleigh branch point. However, a rigorous analytical theory connecting this singular Rayleigh-anomaly behavior to BIC, anapole, as well as to the Q factor scaling, has not been reported. It is worth mentioning that in Ref.~\cite{allayarov2024}, the authors discuss an anapole mechanism for BIC formation in dipole lattices, but their effective dipole polarizability $\alpha_{\mathrm{eff}}$ includes contributions from higher-order multipoles. The resulting lattice anapole relies on destructive interference between different multipole orders, and the authors explicitly state that this mechanism cannot be realized in the pure dipole approximation. Our model shows that a Rayleigh BIC emerges even in a single-dipole approximation when the LR meets the Rayleigh anomaly at the anapole frequency of an isolated metaatom. 

In conclusion, we have predicted a class of BICs in non-subwavelength periodic metasurfaces -- Rayleigh BICs. These states emerge at Rayleigh anomalies, where a collective lattice-resonance branch intersects the individual metaatom anapole frequency. At this singular point, the divergence of the lattice sum is compensated by the divergence of the inverse polarizability, while the anapole condition suppresses radiation into all open diffraction channels. As a result, Rayleigh BICs remain nonradiative even when multiple diffraction channels are open, forming genuine BICs beyond the subwavelength regime. They exhibit an unusual cubic scaling of the Q factor, distinguishing them from symmetry-protected and accidental BICs in metasurfaces. Our results establish a constructive mechanism for engineering nonradiating states beyond the subwavelength regime and connect BIC physics, surface lattice resonances, Rayleigh anomalies, and anapole states within a unified framework.


\textit{Acknowledgments --} We thank Yuri Kivshar, Carsten Rockstuhl, Denis Baranov, Maxim Gorkunov, Yao Liang, Kirill Koshelev, and Mikhail Petrov for useful discussions. The numerical calculations were supported by the Russian Science Foundation (Grant No.~22-11-00153-$\Pi$), and the development of analytical models was supported by the Russian Science Foundation (Grant No.~25-42-10025). This work was also supported by the National Natural Science Foundation of China (Grant No.~W2532010) and the Priority 2030 Academic Leadership Program.





\bibliography{references}

\end{document}